\documentclass[trackchanges, twocolumn]{aastex701}
\newcommand{\LJMU}{\affiliation{Astrophysics Research Institute, Liverpool John Moores University, 146 Brownlow Hill, Liverpool L3 5RF, UK}}

\newcommand{\HUBerlin}{\affiliation{Institut f\"ur Physik, Humboldt-Universit\"at zu Berlin, Newtonstr. 15, 12489 Berlin, Germany}}

\newcommand{\OKC}{\affiliation{Department of Physics, Oskar Klein Centre, Stockholm University, SE-106 91, Stockholm, Sweden}}
\newcommand{\OKCAstro}{\affiliation{Department of Astronomy, Oskar Klein Centre, Stockholm University, SE-106 91 Stockholm, Sweden}}

\newcommand{\Caltech}{\affiliation{Cahill Center for Astronomy and Astrophysics, California Institute of Technology, Mail Code 249-17, Pasadena, CA 91125, USA}}

\newcommand{\IoAKavli}{\affiliation{
Institute of Astronomy and Kavli Institute for Cosmology, University of Cambridge, Madingley Road, Cambridge, CB3 0HA, UK}}

\newcommand{\Birmingham}{\affiliation{School of Physics \& Astronomy and Institute for Gravitational Wave Astronomy, University of Birmingham, Birmingham B15 2TT, UK}}

\newcommand{\INAFRoma}{\affiliation{INAF - Osservatorio Astronomico di Roma, via Frascati 33, 00078 Monte Porzio Catone, Italy}}

\newcommand{\Weizmann}{\affiliation{Department of Particle Physics and Astrophysics, Weizmann Institute of Science, Rehovot, Israel}}

\usepackage{algorithm}
\usepackage{algpseudocode}
\usepackage{gensymb}
\usepackage{amsmath}

\makeatletter
\newcommand{\LongState}[1]{%
  \State \parbox[t]{\dimexpr\linewidth-\ALG@thistlm\relax}{%
    \hangafter=1\hangindent=1.5em\relax#1\strut}%
}
\makeatother

\begin{document}

\title{Follow-up of SN~2025wny VI: The Rate and Detectable Population of Strongly Lensed SLSNe-I in ZTF}



\correspondingauthor{Jacob O. Hjortlund}


\author[0009-0009-6243-8300]{Jacob~O.~Hjortlund}
\OKC
\email[show]{jacob.hjortlund@fysik.su.se}

\author[0000-0002-8380-6143]{Edvard~Mörtsell}
\OKC
\email{edvard@fysik.su.se}


\author[0000-0002-4163-4996]{Ariel~Goobar}
\OKC
\email{ariel@fysik.su.se}

\author[0000-0001-5975-290X]{Joel~Johansson}
\OKC
\email{joeljo@fysik.su.se}

\author[0000-0003-2091-622X]{Avinash~Singh}
\OKCAstro
\email{avinash.singh@astro.su.se}

\author[0000-0001-6343-3362]{Alice~Townsend}
\Birmingham
\email{a.townsend@bham.ac.uk}

\author[0000-0003-3847-0780]{Erin~E.~Hayes}
\IoAKavli
\email{eeh55@cam.ac.uk}

\author[0009-0001-6911-9144]{Maggie L.~Li}
\Caltech
\email{maggieli@caltech.edu}

\author[0000-0001-6797-1889]{Steve~Schulze} 
\Weizmann
\email{steve.schulze@weizmann.ac.il}

\author[0000-0002-2376-6979]{Suhail~Dhawan}
\Birmingham
\email{s.dhawan@bham.ac.uk}

\author[0000-0001-8342-6274]{Jakob~Nordin}
\email{jnordin@physik.hu-berlin.de}
\HUBerlin


\author[0009-0008-2714-2507]{Aleksandra Bochenek}
\LJMU
\email{A.M.Bochenek@2023.ljmu.ac.uk}

\author[0009-0009-9751-9215]{Chiara~Ventura}
\INAFRoma
\email{chiara.ventura@inaf.it}

\author[0000-0002-9646-8710]{Conor~M.~B.~Omand}
\LJMU
\email{C.M.Omand@ljmu.ac.uk}

\author[0000-0003-0733-2916]{Jacob~L.~Wise}
\LJMU
\email{J.L.Wise@2022.ljmu.ac.uk}

\author[0000-0003-1546-6615]{Jesper~Sollerman}
\OKCAstro
\email{jesper@astro.su.se}

\shorttitle{Follow-up of SN~2025wny VI: glSLSNe-I ZTF Rate and Population}
\shortauthors{J. O. Hjortlund et al.}

\begin{abstract}
The Type I superluminous supernova (SLSN-I) SN~2025wny, multiply imaged by
two foreground galaxies at $z_l = 0.375$, is the first confirmed strongly
lensed SLSN and the highest-redshift ($z_s = 2.015$) multiply imaged
supernova in a galaxy-scale configuration. We ask whether one discovery in
seven years of Zwicky Transient Facility (ZTF) operations is consistent with
expectations, and whether SN~2025wny is typical of the detectable
population. We develop a forward simulation of strongly lensed SLSNe-I in
ZTF, combining an empirically calibrated volumetric rate and luminosity
function with a galaxy-scale deflector population, unresolved lensed light
curves, and the actual ZTF observing history. We predict
$0.037^{+0.022}_{-0.020}~{\rm yr}^{-1}$, consistent with the rate of
$0.14^{+0.31}_{-0.11}~{\rm yr}^{-1}$ inferred from a single discovery in seven years of ZTF operations. Conditioning the simulation on discovery by ZTF
corrects the magnifications and intrinsic luminosity of SN~2025wny for
Malmquist- and magnification biases: the debiased peak
bolometric luminosity, $\log_{10}(L_{\rm bol,int}/{\rm erg\,s^{-1}}) =
44.60^{+0.05}_{-0.10}$, places it at the $\sim$97th percentile of the
assumed SLSNe-I luminosity function. SN~2025wny is typical of the detectable population in
redshift, luminosity, and magnification, but its $4.9''$ image separation
exceeds single-galaxy deflector expectations, implying our rates are
conservative for wide-separation systems. 
\end{abstract}

\keywords{\uat{Supernovae}{1668} --- \uat{Strong gravitational lensing}{1643} --- \uat{Observational cosmology}{1146} --- \uat{Time domain astronomy}{2109}}

\section{Introduction}
\label{sec:introduction}

Strong gravitational lensing of supernovae (SNe) by foreground galaxies
and galaxy clusters provides two complementary scientific opportunities.
First, the arrival-time differences between multiple images are directly
sensitive to the expansion rate of the Universe, enabling measurements of
the Hubble constant, $H_0$, that are independent of both the local
distance ladder and early-Universe physics
(\citealp{refsdal_possibility_1964}, see \citealp{birrer_time-delay_2024} for a recent review); such measurements have gained renewed
importance in light of the persistent tension between early- and
late-Universe determinations of $H_0$ (\citealp{di_valentino_realm_2021}, and references therein). Second, the lensing magnification acts as a gravitational
telescope, boosting the observed flux of high-redshift sources and
granting observational access to transient populations that would
otherwise fall below survey detection limits \citep{suyu_strong_2024}.

Despite these prospects, strongly lensed SNe (glSNe) remain rare. Since
the discovery of the core-collapse supernova SN Refsdal
\citep{kelly_multiple_2015, kelly_sn_2016}, the first multiply imaged supernova
lensed by a galaxy cluster, the number of known glSNe has increased
rapidly. These discoveries include cluster-scale events
\citep{rodney_gravitationally_2021, chen_shock_2022, pierel_lensed_2024, coulter_spectroscopically_2026, dhanasingham_sn_2026}, as well as galaxy-scale lensed Type Ia supernovae discovered in wide-field time-domain surveys \citep{goobar_iptf16geu_2017, goobar_uncovering_2023}; see \cite{goobar_strongly_2025} for a recent review. More recently,
two strongly lensed supernovae originating from explosions of massive
stars have been identified: SN~2025mkn \citep{lemon_natural_2026} and the subject
of this paper series, the Type I superluminous supernova (SLSN-I)
SN~2025wny \citep{johansson_discovery_2025,taubenberger_holismokes_2026}.

\begin{figure*}[ht!]
\plotone{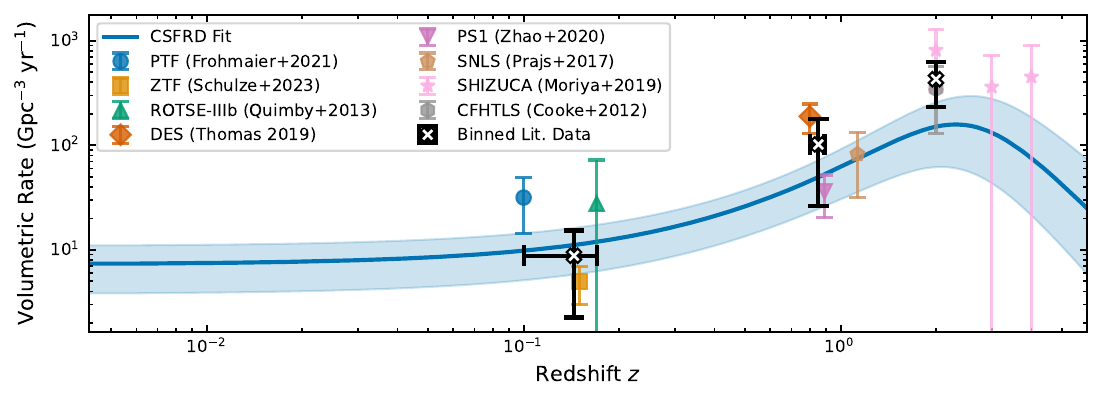}
\caption{
Best-fit model of the volumetric rate of SLSNe-I fit to literature data.
The solid blue line shows the posterior median, and the blue band the
$68\%$ confidence interval. Colored markers show the individual literature
measurements, as labeled in the legend. Sets of measurements that exhibit
systematic disagreement have been binned and had their uncertainties increasedto account for systematics, shown as
black-and-white cross markers: (PTF, ZTF, ROTSE-IIIb), (DES, PS1), and
(SHIZUCA, CFHTLS), respectively.
\label{fig:slsne_rates}
}
\end{figure*}

The scarcity of discovered systems has motivated considerable effort to
forecast glSNe yields for wide-field surveys, both to guide search
strategies and to assess the cosmological potential of future samples
\citep{oguri_gravitationally_2010, goldstein_rates_2019, wojtak_magnified_2019, carracedo_detectability_2024, arendse_detecting_2024, sainz_de_murieta_find_2024, abe_halo_2025}. These forecasts have concentrated
primarily on SNe Ia, whose standardizable luminosities and
well-characterized volumetric rates make predictions comparatively
robust, and on the aggregate core-collapse population. SLSNe-I have
received far less attention, despite properties that make them
distinctive gravitationally lensed sources: their extreme peak
luminosities ($M \lesssim -21$~mag; \citealp{quimby_hydrogen-poor_2011}) allow magnified events to be detected at
redshifts beyond the reach of other SN types, and their long
light-curve durations, further stretched by cosmological time dilation, relax the follow-up cadence required for time-delay measurements. At the
same time, SLSNe-I are intrinsically rare, with volumetric rates of
$\sim 10^{-4}$--$10^{-3}$ times that of SNe Ia
\citep{quimby_rates_2013, prajs_volumetric_2017, frohmaier_core_2021, perley_zwicky_2020},
they exhibit strong preferences for low-mass, low-metallicity host
galaxies \citep{lunnan_hydrogen-poor_2014, perley_host-galaxy_2016, schulze_cosmic_2018}, and their powering mechanism, commonly modeled as
magnetar spin-down \citep{kasen_supernova_2010, woosley_bright_2010, nicholl_magnetar_2017}, remains debated \citep{angus_superluminous_2019, chen_hydrogen-poor_2023}. Consequently, both the volumetric rate
and the luminosity function of SLSNe-I are poorly constrained,
particularly at $z>1$, and forecasts of lensed-SLSN yields
inherit these uncertainties.

SN~2025wny provides the first observational anchor for such forecasts.
The transient was first detected by the Zwicky Transient Facility
\citep{bellm_zwicky_2019, graham_zwicky_2019} on 2025 August 23
and independently reported by the Gravitational-wave Optical Transient
Observer \citep{steeghs_gravitational-wave_2022}, and was subsequently classified as an SLSN-I at
$z_s = 2.015$, multiply imaged by two foreground galaxies at
$z_l = 0.375$ associated with the known lens candidate PS1~J0716+3821
\citep{johansson_discovery_2025, taubenberger_holismokes_2026, canameras_holismokes_2020, collaboration_data_2026}.
Follow-up imaging resolved five SN images with a maximum image
separation of $4.9''$ \citep{Goobar2026}. SN~2025wny is the first confirmed strongly lensed SLSN, the highest-redshift multiply imaged SN identified in a galaxy-scale lensing configuration to date, and one of the best-observed
high-redshift SLSNe of any kind. This paper is the sixth in a series
presenting the follow-up campaign of SN~2025wny: \cite{Goobar2026} present
the HST and JWST imaging and spectroscopy; \cite{Li2026} analyze the SN
physics at cosmic noon; \cite{Johansson2026} measure spectroscopic time delays
between the images; \cite{Townsend2026} present photometric time-delay
measurements; \cite{Mortsell2026} present the lens modelling and the resulting
$H_0$ inference; and \cite{Qin2026} study the host galaxy.

In this work we address the population-level questions raised by the
discovery. Is the detection of one glSLSN-I in seven years of ZTF
operations consistent with expectations given the known SLSN-I
population and the ZTF selection function? And is SN~2025wny itself,
with its high apparent brightness and large magnification, a typical representative of the ZTF-detectable
glSLSNe-I population? To answer these questions, we develop a forward
simulation of strongly lensed SLSNe-I in ZTF, combining an empirically
calibrated SLSN-I volumetric rate and luminosity function with a
galaxy-scale deflector population, realistic unresolved light curves,
and the actual ZTF observing history, and we convert the simulated
catalog into physical detection rates via importance sampling. Beyond
rate predictions, the same simulation framework enables an event-level
application: by conditioning the simulated population on discovery by
ZTF, we correct the inferred magnification and intrinsic peak
luminosity of SN~2025wny for the Malmquist- and magnification-type
selection biases that affect any magnitude-limited lensed-transient
discovery, providing debiased posteriors that complement the lens-model
constraints of \cite{Mortsell2026}.

The paper is structured as follows. Section~\ref{sec:simulations}
describes the simulation framework: the SLSN-I source population
(Sec.~\ref{sec:slsne_population}), the deflector population
(Sec.~\ref{sec:deflector_population}), the construction of unresolved
lensed light curves (Sec.~\ref{sec:unresolved_lightcurves}), the ZTF
survey model (Sec.~\ref{sec:ztf_survey}), the simulated glSLSNe-I
catalog (Sec.~\ref{sec:glslsne_simulation}), and the
importance-sampling weights used to recover physical rates
(Sec.~\ref{sec:importance_sampling}). We present the predicted
detection rates for ZTF and compare them with the
observed rate in Section~\ref{sec:detection_rates}, apply the debiasing
framework to SN~2025wny in Section~\ref{sec:debiasing}, and compare
SN~2025wny to the predicted properties of the ZTF-detectable glSLSNe-I
population in Section~\ref{sec:population_properties}. We summarize our
conclusions in Section~\ref{sec:summary}.

\section{Simulation Framework}
\label{sec:simulations}

We estimate the observer-frame rate of ZTF-detectable strongly lensed
SLSNe-I (glSLSNe-I) using a forward simulation combined with importance
sampling. The calculation separates the problem into two parts. First, we
generate a large catalog of accepted strong-lensing configurations and
simulate the unresolved glSLSNe-I light curves that would be observed by
ZTF. Second, we assign each simulated system an importance weight that
converts it from the proposal distributions used in the simulation into its
contribution to the physical glSLSNe-I detection rate.

Schematically, the target detection rate is written as
\begin{align}
    \dot{N}_{\rm det}
    =
    \frac{\Omega_{\rm survey}}{4\pi}
    \int_0^{z_{\rm max}}
    dz_s\,
    \frac{r_{\rm SLSN}(z_s)}{1+z_s}
    \left.\frac{dV_{\rm C}}{dz}\right\rvert_{z=z_s}
    P_{\rm gl,det}(z_s),
    \label{eq:target_detection_rate}
\end{align}
where $r_{\rm SLSN}(z_s)$ is the rest-frame volumetric rate of SLSNe-I,
the factor $(1+z_s)^{-1}$ converts the rate to the observer frame,
$dV_{\rm C}/dz$ is the full-sky comoving volume element per unit
redshift, and $\Omega_{\rm survey}$ is the effective survey solid angle.
The term $P_{\rm gl,det}(z_s)$ is the probability that a source at redshift $z_s$ is both strongly lensed and detected by the survey. Throughout this work, we assume a standard flat $\Lambda$CDM model with $H_0 = 67.66 \, {\rm km} \, {\rm s}^{-1} \, {\rm Mpc}^{-1}$ and $\Omega_m = 0.30966$ \citep{planck_collaboration_planck_2020}.

For a given source redshift, this probability depends on the foreground
deflector population, the strong-lensing cross-section, the unresolved
lensed light curve, the ZTF observing cadence, and the photometric and
spectroscopic selection criteria. Denoting the deflector parameters by
$\boldsymbol{\theta}_{\rm sl} = (z_l, \sigma, q, \phi_l, \gamma,
\phi_\gamma)$ (Sec.~\ref{sec:deflector_population}), we can write this
dependence schematically as
\begin{align}
    P_{\rm gl,det}(z_s)
    =
    &\int_0^{z_s} dz_l\,
    \frac{dV_{\rm C}}{dz_l\, d\Omega}
    \int d\sigma\,dq\,d\gamma\,
    \frac{dn}{d\sigma} \nonumber \\
    &p(q \mid \sigma)\,
    p(\gamma)\,
    \tau_{\rm sl}(\boldsymbol{\theta}_{\rm sl}, z_s)\,
    \epsilon_{\rm det},
    \label{eq:lensed_detection_probability}
\end{align}
where $z_l$ is the lens redshift, $\sigma$ the lens velocity
dispersion, $q$ the projected axis ratio, $\gamma$ the external shear
magnitude, $\tau_{\rm sl}$ the angular strong-lensing cross-section,
i.e.\ the source-plane solid angle over which a source produces multiple
images, and $\epsilon_{\rm det}$ the probability that the resulting
unresolved glSLSN-I satisfies our ZTF photometric and spectroscopic
selection criteria (Sec.~\ref{sec:ztf_survey}). The position angles
$\phi_l$ and $\phi_\gamma$ are sampled uniformly and are suppressed
here for clarity.

In practice, evaluating Equations~\ref{eq:target_detection_rate} and
\ref{eq:lensed_detection_probability} directly is inefficient, because the
strong-lensing cross-section is small and the survey selection function
depends on the detailed light-curve realization and ZTF observing history.
We therefore draw simulated systems from proposal distributions chosen for
computational efficiency, require that they produce multiple images, inject
the resulting unresolved light curves into the ZTF survey simulation, and
recover the physical rate using importance weights.

The remainder of this section describes each component of the calculation.
Section~\ref{sec:slsne_population} defines the SLSNe-I source population and
volumetric-rate model. Section~\ref{sec:deflector_population} describes the
foreground deflector population. Section~\ref{sec:unresolved_lightcurves}
describes how the individual lensed images are combined into unresolved
observed light curves. Section~\ref{sec:ztf_survey} describes the ZTF survey
simulation and detection criteria. Section~\ref{sec:glslsne_simulation}
describes how the simulated catalog of glSLSNe-I is generated, and
Section~\ref{sec:importance_sampling} gives the importance weights used to
convert the simulated catalog into a physical rate.

\subsection{SLSNe-I Source Population}
\label{sec:slsne_population}

SLSNe-I exhibit a notably low intrinsic volumetric rate compared to other
transient classes \citep{frohmaier_core_2021}. Consequently, the volumetric
rate of SLSNe-I remains poorly constrained at higher redshifts, where
detections are sparse. Figure~\ref{fig:slsne_rates} shows volumetric rate
measurements collected from the literature
\citep{schulze_1100_2024,frohmaier_core_2021,zhao_event_2021,moriya_first_2019,prajs_volumetric_2017,quimby_rates_2013,cooke_superluminous_2012, thomas_volumetric_2019}.

Although the progenitor channels of SLSNe-I are not well understood, all
proposed channels fundamentally tie the volumetric rate of SLSNe-I to the
cosmic star formation rate density (CSFRD). However, SLSNe-I demonstrate
strong environmental preferences: they are observed to occur more
frequently in low-mass and low-metallicity galaxies, with low metallicity
potentially serving as the stronger indicator of suitable progenitor
environments \citep{lunnan_hydrogen-poor_2014, perley_host-galaxy_2016, schulze_cosmic_2018}.

To account for these environmental dependencies, we parametrize the
SLSNe-I volumetric rate as proportional to the CSFRD modulated by a
logistic function. This parametrization allows for an up-weighting of the
volumetric rate at higher redshifts, tracking the expected cosmic
metallicity evolution. We define the SLSNe-I volumetric rate density as
\begin{align}
    r_{\rm SLSN}(z) = A \frac{\left( 1+z \right)^a}{
        1 + \left[ \left( 1+z\right) / B\right]^b
    } \frac{1}{1+\exp\left[ -k (z-z_0) \right]}.
    \label{eq:slsne_rate}
\end{align}
We fit Eq.~\ref{eq:slsne_rate} to the literature volumetric rate
measurements using \textsc{emcee} \citep{foreman-mackey_emcee_2013}. We
consider two models, with $k$ and $z_0$ either free or fixed to
$k=z_0=0$; in the latter case the logistic term reduces to a constant
factor of $1/2$, which is absorbed into $A$. We adopt priors on the CSFRD
parameters from \citet{fujimoto_alma_2024}. To perform the fit, we bin
sets of measurements that exhibit clear systematic disagreement: (PTF,
ZTF, ROTSE-IIIb) \citep{frohmaier_core_2021, schulze_1100_2024,
quimby_rates_2013}, (DES, PS1) 
\citep{thomas_volumetric_2019,zhao_event_2021}, and (SHIZUCA, CFHTLS)
\citep{moriya_first_2019, cooke_superluminous_2012}. The measurements are combined using an uncertainty-weighted mean, and a systematic uncertainty is added in quadrature to ensure a reduced-$\chi^2=1$. The binned
measurements are shown as white-cross markers in
Figure~\ref{fig:slsne_rates}.

Allowing the logistic parameters $k$ and $z_0$ to vary yields a negligible improvement over the four-parameter model, with $\ln\mathcal L_{\max,6}-\ln\mathcal L_{\max,4}\simeq0.02$. Given the six volumetric-rate measurements available, we therefore adopt the simpler model, with $k=z_0=0$ and the constant logistic factor absorbed into $A$, as our fiducial parametrization. The resulting best-fit parameters for this model are
$A=14\pm7$ Gpc$^{-3}$ yr$^{-1}$, $B=3.4\pm0.3$, $a=3.1\pm0.5$, $b=6.6\pm0.5$. The posterior-median
volumetric rate, alongside the 16th--84th percentile confidence interval,
is shown in Figure~\ref{fig:slsne_rates}. 

\begin{figure}
    \centering
    \includegraphics[width=\linewidth]{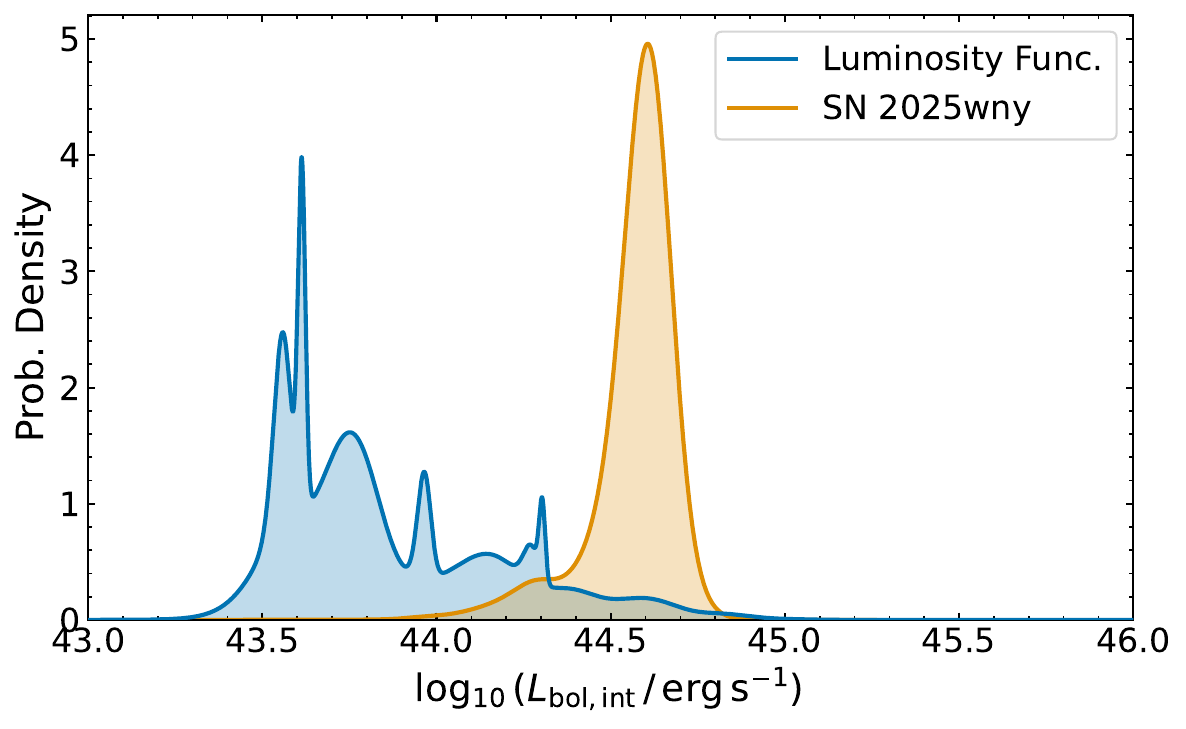}
    \caption{Intrinsic peak bolometric luminosity function of SLSNe-I
    (blue), constructed from the spectroscopically complete $m_{\rm peak} < 19$ subsample of
    \citet{chen_hydrogen-poor_2023} via Eq.~\ref{eq:luminosity_function}.
    For comparison, the orange curve shows the debiased posterior on the
    intrinsic peak bolometric luminosity of SN~2025wny derived in
    Sec.~\ref{sec:debiasing}.}
    \label{fig:luminosity}
\end{figure}

We make use of SLSNe-I spectral templates from \citet{kessler_models_2019}.
These templates were generated by sampling model parameters from a
multivariate Gaussian distribution obtained by fitting the \textsc{MOSFiT}
\texttt{slsn} model, which assumes a magnetar central engine and a
blackbody SED with a linear suppression for $\lambda < 3000$~\AA, to 58
well-observed SLSNe-I \citep{nicholl_magnetar_2017, villar_theoretical_2017}. It has been
suggested that the magnetar model cannot capture the full diversity of
observed SLSNe-I \citep{angus_superluminous_2019, chen_hydrogen-poor_2023}. To increase the diversity of the simulated population, we therefore
decouple the template shape from the intrinsic peak bolometric luminosity,
$L_{\rm bol,int}$: the templates provide the SED shape and time evolution,
while each simulated SLSN-I is rescaled to a value of $L_{\rm bol,int}$
drawn from an assumed luminosity function. 

The luminosity function of SLSNe-I is not well constrained, with most
estimates being limited to small sample sizes and low redshift. For the
purpose of this work we construct an empirical luminosity function from
the ZTF SLSNe-I sample of \citet{chen_hydrogen-poor_2023} by applying a
peak apparent magnitude cut of $m_{\rm peak} < 19$, the limit down to
which the parent catalog is assessed to be spectroscopically complete.
The selection function of the resulting subsample of $N=33$ SLSNe-I
therefore reduces to the sharp magnitude cut itself. Each SN in the
subsample is weighted by the inverse of the comoving volume over which
it could have entered the sample,
$V_{{\rm max},i} = V_{\rm C}\left(z_{{\rm max},i}\right)$, where
$z_{{\rm max},i}$ is the redshift at which its peak apparent magnitude
would equal the cut. The luminosity function is then a Gaussian mixture
over the observed intrinsic peak bolometric luminosities,
\begin{align}
    p\left(L_{\rm bol,int}\right)
    \propto
    \sum_i^{N} \frac{1}{V_{{\rm max},i}}
    \mathcal{N}\left( L_{\rm bol,int} \mid
    L_{{\rm bol,int},i}, \sigma_{L,i} \right),
    \label{eq:luminosity_function}
\end{align}
where $\sigma_{L,i}$ is the measurement uncertainty on the $i$-th peak
luminosity. The resulting luminosity function is shown in
Figure~\ref{fig:luminosity}. The luminosity function shows distinct
features, owing to the small effective sample size in addition to some
SLSNe-I having small uncertainties on their peak luminosities.

\subsection{Deflector Population}
\label{sec:deflector_population}

We model the simulated lens galaxies as Singular Isothermal Ellipsoids
\citep[SIE;][]{kormann_isothermal_1994}, which provide a good approximation
to the total mass profiles of elliptical galaxies. We additionally include
an external shear term to account for the impact of nearby and
line-of-sight structures \citep{wong_effect_2010}. The resulting deflector
model is parametrized by
$\boldsymbol{\theta}_{\rm sl} = (z_l, \sigma, q, \phi_l, \gamma,
\phi_\gamma)$: the lens redshift, the velocity dispersion, the projected
axis ratio, the lens position angle, the external shear magnitude, and the
external shear position angle.

We assume that lens galaxies are distributed uniformly in comoving volume,
\begin{align}
    p(z_l) \propto \left. \frac{d V_{\rm C}}{d z}\right\rvert_{z=z_l},
    \label{eq:zl}
\end{align}
where $dV_{\rm C}/dz$ is the comoving volume element per unit redshift.
When sampling, we condition on the source redshift $z_s$, such that
$p(z_l)$ is truncated to $z_l\in[0, z_s]$.

The velocity dispersion distribution is modeled as a modified Schechter
function \citep{sheth_velocity_2003},
\begin{align}
    dn = \phi_* \left( \frac{\sigma}{\sigma_*} \right)^\alpha \exp\left[ -\left( \frac{\sigma}{\sigma_*} \right)^\beta \right] \frac{\beta}{\Gamma(\alpha/\beta)} \frac{d\sigma}{\sigma},
    \label{eq:vd}
\end{align}
where $dn$ is the differential number of lenses per unit comoving volume.
We adopt the parameter values derived by \citet{bernardi_galaxy_2010} from
SDSS DR6 for all galaxy types, $\phi_*=2.099 \times 10^{-2} (h/0.7)^3$
Mpc$^{-3}$, $\sigma_* = 113.78$ km s$^{-1}$, $\alpha=0.94$, and
$\beta=1.85$. We assume no redshift evolution in the velocity dispersion
distribution and sample $\sigma$ according to Eq.~\ref{eq:vd} on the
truncated range $\sigma \in [50, 400]$ km s$^{-1}$.

We follow \citet{collett_population_2015} and draw the axis ratio $q$
conditional on the velocity dispersion $\sigma$ from a Rayleigh
distribution in the ellipticity $1-q$,
\begin{align}
    p(1-q \mid \sigma) = \frac{1-q}{s^2}\exp \left[ -\frac{1}{2} \left( \frac{1-q}{s} \right)^2 \right],
    \label{eq:ellip}
\end{align}
where $s=A_q+B_q\sigma$, with $A_q=0.38$ and
$B_q=5.7\times 10^{-4} \, ({\rm km\,s^{-1}})^{-1}$. The conditional
Rayleigh distribution encodes the tendency of more massive galaxies to be
closer to spherical. We truncate sampling of $q$ to the range
$q\in[0.2, 1.0]$ to exclude highly flattened mass profiles.

We assume a Rayleigh distribution for the external shear magnitude, with
scale $s_\gamma=0.05$ \citep{wong_effect_2010},
\begin{align}
    p(\gamma) = \frac{\gamma}{s_\gamma^2} \exp\left(-\frac{\gamma^2}{2s_\gamma^2}\right).
    \label{eq:shear}
\end{align}

The position angles are sampled uniformly, with
$\phi_l \sim U[0, 2\pi]$ and $\phi_\gamma\sim U[0, 2\pi]$.

\subsection{Unresolved glSLSNe-I Light Curves}
\label{sec:unresolved_lightcurves}

In practice, most glSLSNe-I will appear unresolved to ground-based
wide-field surveys, as typical image separations are smaller than the spatial resolutions of such surveys.
In most cases the individual images are therefore not distinguishable, and
the system appears as a single object combining the flux from all
individual images.

Given an accepted strong-lensing configuration
(Sec.~\ref{sec:deflector_population}) and an SLSN-I source
(Sec.~\ref{sec:slsne_population}), we apply the magnifications and time
delays of the individual images, such that the combined unresolved
glSLSN-I light curve is given by
\begin{align}
    F_{\rm glSLSN}(z_s, t)
    =
    \sum_{i=1}^{N_{\rm img}} \mu_i\, F_{\rm SLSN}(z_s, t-\Delta t_i),
    \label{eq:unresolved_flux}
\end{align}
where $F_{\rm SLSN}(z_s, t)$ is the flux derived from the spectral
template, dependent on the source redshift and the observer-frame time,
$\mu_i$ and $\Delta t_i$ are the magnification and time delay of the
$i$-th image, with delays measured relative to the first-arriving image
(i.e.\ $\Delta t_1 = 0$), and $N_{\rm img}$ is the total number of images.

We model the effects of reddening due to Milky Way extinction as well as
attenuation due to the intergalactic medium. For Milky Way extinction, we evaluate extinction at the sky positions assigned in Sec.~\ref{sec:glslsne_simulation} using the Galactic dust maps of \citet{planck_collaboration_planck_2014}. We neglect reddening due to host-galaxy dust, since SLSNe-I are typically found in environments with low dust-extinction \citep{lunnan_hydrogen-poor_2014, perley_host-galaxy_2016, schulze_cosmic_2018}. Similarly, the dust
properties of lens galaxies are not well constrained, but the small sample
of observed systems indicates that lens extinction is small \citep{goobar_uncovering_2023, goobar_iptf16geu_2017}. For attenuation due to intergalactic neutral hydrogen, we use the mean attenuation curves of \citet{inoue_updated_2014}, calibrated against absorption measurements out to $z\sim6$, encompassing our simulation limit of $z_{\max}=5.5$. This correction becomes increasingly important at high source redshift as absorption blueward of rest-frame Lyman-$\alpha$ enters the observed optical bands. In ZTF $g$-band, absorption begins to affect the blue edge at $z_s\sim2.4$, reaching 464 nm at $z_s\simeq2.8$, while the redder bands are affected at progressively higher redshifts. For illustration, at $z_s=3$ the prescription predicts a mean transmission of approximately $0.71$ at $464$ nm, equivalent to $0.37$ mag of monochromatic attenuation; the band-integrated correction depends on the source spectrum and filter response.

While the effects of microlensing by individual stars can add noise to
glSNe light curves \citep{dobler_microlensing_2006}, this has been found to
have a modest impact on population-level glSNe yields
\citep{arendse_detecting_2024}. We neglect the effects of
microlensing when simulating glSLSNe-I and return to this choice in
Sec.~\ref{sec:population_properties}.

\subsection{ZTF Survey Observation Logs}
\label{sec:ztf_survey}

The Zwicky Transient Facility (ZTF) uses a camera mounted on the Samuel
Oschin Telescope at the Palomar Observatory in California to scan the night
sky for transient and variable objects \citep{bellm_zwicky_2019}. ZTF is
led by the California Institute of Technology (Caltech) in collaboration
with partners worldwide. It is an effective facility for discovering new
transients due to its wide field of view of $47 \, \text{deg}^2$,
relatively high cadence of $1$--$3$ days, and median $5\sigma$ depth of
$20.5$~mag for $30$~s exposures \citep{dekany_zwicky_2020}.

ZTF observes in three filters ($g$, $r$, and $i$), mapping the northern sky
down to a minimum declination of $-30^{\circ}$. The public $g$ and $r$ bands are
used most frequently, while the proprietary $i$ band covers a smaller region of the
sky. The ZTF collaboration maintains multiple surveys, among them the
Bright Transient Survey (BTS, \citealp{fremling_zwicky_2019})  and the ZTF + LSST Inventory of
Nuclear Transients and Supernovae (Z+LIONS, \citealp{hinds_ztflsst_2026}) survey. BTS makes use of a low-resolution
integral-field spectrograph, SEDm \citep{blagorodnova_sed_2018,rigault_fully_2019}, to construct a
magnitude-limited, spectroscopically classified SN sample. The survey has
an effective sky area of $\Omega_{\rm BTS} = 14{,}400 \, \text{deg}^2$
\citep{perley_zwicky_2020} and aims to be spectroscopically complete for
peak magnitudes $m_{\rm peak} < 18.5$ in all three ZTF filters. The spectroscopic selection function
of BTS is well modeled by a sigmoid, as shown in Sec.~4.6 of
\citet{rigault_ztf_2025},
\begin{align}
    \epsilon_{\rm det}(m_{\rm peak}) = \left( 1 + e^{s(m_{\rm peak}-m_0)} \right)^{-1}, \label{eq:det_eff}
\end{align}
where $m_0 = 18.8$ and $s = 4.5$, such that $\epsilon_{\rm det}\to 1$ for
bright events and reaches $50\%$ at $m_{\rm peak} = m_0$.

The Z+LIONS survey is a magnitude-limited spectroscopic survey of ZTF
transients in the Vera C. Rubin Observatory's Legacy Survey of Space and Time (Rubin LSST) Wide-Fast-Deep footprint \citep{ivezic_lsst_2019},
targeting $>95\%$ spectroscopic completeness for peak magnitudes
$m_{\rm peak} < 20$. The survey covers
$\Omega_{\rm Z+LIONS} = 10{,}000 \, \text{deg}^2$ of the Rubin footprint
with 1-day cadence and 30~s exposures in the ZTF $g$, $r$, and $i$ bands,
with transients identified by filtering the ZTF and LSST alert streams.

For this work we consider observations taken under the ZTF public survey,
as well as initially proprietary partnership time, covering 2.7~yr of the
survey from June 19, 2018 to February 28, 2021.
The typical exposure time is 30~s and the typical cadence is $\sim 3$
days; both vary across filters, time, and sky position, mainly due to
atmospheric conditions, season, exposure length, and detector sensitivity.
When computing detection rates (Sec.~\ref{sec:detection_rates}), we adopt
the effective BTS sky area for the fiducial and BTS selection functions.
For Z+LIONS, we adopt the sky area defined by the ZTF--Rubin footprint
overlap and set $m_0=20.6$ in Eq.~\ref{eq:det_eff} (with $s=4.5$),
corresponding to the Z+LIONS completeness goal. Finally, owing to
limitations of the available observation logs, we do not model the increase
in cadence from $\sim 3$ days to $1$ day under Z+LIONS; the corresponding
rates in Sec.~\ref{sec:detection_rates} are therefore conservative in this
respect.

The spectroscopic selection models of BTS and Z+LIONS do not explicitly reproduce the candidate-vetting procedure used in the ZTF search for glSNe. Candidates are primarily vetted manually by comparing their inferred absolute magnitude at the spectroscopic or photometric redshift of a potential deflector, where available, with that expected for an unlensed SN~Ia. Cross-matching against catalogues of known lenses and lens candidates provides additional support for the lensing interpretation, but is not a prerequisite for selection. For SN~2025wny, flagging relied on the known spectroscopic redshift of the potential deflector, while its prior identification as a lens candidate, cross-matched through the Strong Lensing Database (SLED\footnote{https://sled.amnh.org/}), provided additional confidence \citep{johansson_discovery_2025}.

We describe these selection stages through a candidate-flagging efficiency $\epsilon_{\rm flag}$ and a spectroscopic classification
efficiency $\epsilon_{\rm spec}(m_{\rm peak})$. Here, $\epsilon_{\rm flag}$ depends on the availability and reliability
of redshift estimates for potential deflectors, the association of transients with those galaxies, and the criteria used to identify
anomalously luminous events. We model only the spectroscopic efficiency, through Eq.~\ref{eq:det_eff}, and adopt $\epsilon_{\rm flag}=1$, thereby neglecting incompleteness at the candidate-flagging stage. Quantifying this incompleteness would require characterising the relevant galaxy and redshift information together with the candidate-selection procedure across the survey footprint; it cannot be robustly calibrated from a single discovery. By contrast, requiring a classifying spectrum is conservative with respect to candidate identification: a transient associated with a potential deflector of known redshift could in principle be recognised as a lensing candidate from its excess brightness alone, even if it is too faint for spectroscopic classification.

\subsection{Simulation of Strongly Lensed SLSNe-I}
\label{sec:glslsne_simulation}

We simulate a catalog of glSLSNe-I by first generating valid strong-lensing
configurations and then assigning SLSNe-I light curves, sky positions,
explosion times, and survey observations to each configuration. This
procedure is designed to efficiently sample the rare geometric
configurations that produce multiple imaging, while the physical rate
associated with each accepted system is accounted for through the
importance weights described in Section~\ref{sec:importance_sampling}.

We generate $N_{\rm sl}=10^7$ accepted strong-lensing systems.
Source redshifts are drawn uniformly over $z_s \sim U(z_{\rm min},z_{\rm max})$,
with $z_{\rm min}=0.1$ and $z_{\rm max}=5.5$. This redshift limit was
chosen by simulating SLSNe-I magnified by $\mu=500$ uniformly in redshift
according to Sec.~\ref{sec:slsne_population}, and finding the redshift at
which the ZTF $r$-band $5\sigma$ depth, $m_{\rm lim}=20.5$,
intersects the bright $3\sigma$ tail.
This intersection occurs at $z_{3\sigma} = 5.35$, which we round up to
$z_{\rm max}=5.5$. For each source redshift, we draw a lens galaxy with
parameters $\boldsymbol{\theta}_{\rm sl}$ from the deflector population
described in Section~\ref{sec:deflector_population}.

For each proposed lens configuration, we identify the region of the source
plane that can produce multiple images. To do this efficiently, source
positions $(x_s,y_s)$ are drawn uniformly from within the minimum
bounding box of the lens caustics, the area of which is denoted
$A_{\rm box}$. We then evaluate the lensing configuration and accept the
system if it produces $N_{\rm img} \geq 2 $. The number of lens configuration draws required to obtain the $i$-th
accepted system, $N_{{\rm draw},i}$, is stored for use in the
importance-weight calculation.

For each accepted strong-lensing configuration, we simulate
$N_{\rm SLSN}=100$ independent SLSN-I realizations. Their intrinsic light
curves and peak luminosities are drawn from the source model described in
Section~\ref{sec:slsne_population}. The simulated SNe are assigned sky
positions drawn uniformly in solid angle over the ZTF survey footprint,
with right ascension and declination given by
\begin{align}
    \alpha_{\rm sky} \sim U(0^\circ,360^\circ),
    \qquad
    \sin\delta_{\rm sky} \sim U\left(\sin(-30^\circ), 1\right),
\end{align}
such that $\delta_{\rm sky} \in [-30^\circ, 90^\circ]$.
We also draw the explosion time $t_0$ uniformly over the time span
covered by the ZTF survey observations. In total, the simulation therefore
produces  $N_{\rm glSLSN}= N_{\rm sl}N_{\rm SLSN}=10^9$ lensed SLSN-I realizations.

The lensed images of each SLSN-I are combined into an unresolved observed
light curve as described in Section~\ref{sec:unresolved_lightcurves}. The
unresolved light curves are then processed through the ZTF survey logs
using the Python package \texttt{skysurvey} \citep{rigault_skysurvey_2026}, giving simulated observations
with realistic cadence, filter coverage, limiting magnitudes, and
observational uncertainties.

We define a photometric detection using the simulated ZTF observations. A
single observation is counted as a detection if it has flux
signal-to-noise ratio ${\rm SNR} > 5 .$
Although ZTF alert production requires at least two detections, an object
with only two detections over its full light-curve evolution would
generally be insufficient for reliable follow-up and classification. We
therefore impose a stricter detection requirement: a simulated glSLSN-I
must have at least five detections within $t_{\rm det} \in [-10,+20]~{\rm days}$
of the observed light-curve peak $t_{\rm peak}$. These detections must span at least five days in observer-frame time. This last cut mainly impacts high-redshift transients, where it ensures that the light-curve spans ~1 rest-frame day, since classification is not typically possible purely based on short-term variation. The chosen cuts mean that the resulting rates are conservative, especially in the high-redshift tail.

Finally, we record the quantities needed to model spectroscopic selection.
In particular, for each simulated realization we store the peak apparent
magnitude of the unresolved lensed light curve, $m_{\rm peak}$. These peak
magnitudes are used in Section~\ref{sec:importance_sampling} to compute the
system-level spectroscopic-selection efficiency. The resulting simulated
catalog therefore contains, for each accepted strong-lensing system, the
lensing configuration, the effective source-plane sampling area, the number
of source-position trials, the simulated ZTF detections, and the quantities
needed to assign the physical importance weight.

\subsection{Importance-Sampling Weights}
\label{sec:importance_sampling}

The simulations described in Section~\ref{sec:glslsne_simulation} generate
accepted strong-lensing configurations from proposal distributions rather
than directly from the physical event-rate distribution. We therefore
assign each accepted simulated system an importance weight that converts it
into its contribution to the physical rate of detectable glSLSNe-I.

The simulation pipeline described in the previous sections can be written
as a sampling distribution
\begin{align}
    p&_{\rm samp}(z_s, \boldsymbol{\theta}_{\rm sl}) = p(z_s) p\left(
    \boldsymbol{\theta}_{\rm sl} \mid N_{\rm img} \geq 2, z_s
    \right) \nonumber \\
    & = p(z_s) \frac{
    p\left(
    N_{\rm img} \geq 2 \mid \boldsymbol{\theta}_{\rm sl}, z_s
    \right) p(\boldsymbol{\theta}_{\rm sl} \mid z_s)
    }{p(N_{\rm img} \geq 2 \mid z_s)} \nonumber \\
    & =
    \frac{1}{z_{\rm max}-z_{\rm min}}
    \frac{\tau_{\rm sl}(\boldsymbol{\theta}_{\rm sl}, z_s)}{A_{\rm box}(\boldsymbol{\theta}_{\rm sl}, z_s)}
    \frac{
    p(\boldsymbol{\theta}_{\rm sl} \mid z_s)
    }{p(N_{\rm img} \geq 2 \mid z_s)}.
    \label{eq:sampling_dist}
\end{align}
The second equality follows from Bayes' theorem. The third follows from
the uniform source-redshift proposal and from the fact that, for source
positions drawn uniformly within the caustic bounding box, the probability
of obtaining a valid strong-lensing configuration is
$\tau_{\rm sl}/A_{\rm box}$. The distribution
$p(\boldsymbol{\theta}_{\rm sl} \mid z_s)$ is the proposal distribution
over the deflector parameters,
\begin{align}
    &p(\boldsymbol{\theta}_{\rm sl} \mid z_s) = \nonumber \\
    &\frac{p(z_l)}{V_{\rm C}(z_s)}
    \frac{1}{N_{\rm gal}}\frac{dn}{d\sigma}
    \frac{p(1-q \mid \sigma)}{N_q}
    \frac{p(\gamma)}{N_\gamma}
    p(\phi_l) p(\phi_\gamma),
    \label{eq:lens_prior}
\end{align}
where $V_{\rm C}(z_s)$, $N_{\rm gal}$, $N_q$, and $N_\gamma$ are the
normalizations corresponding to the truncated sampling ranges described in
Sec.~\ref{sec:deflector_population}. Finally, the acceptance probability
$p(N_{\rm img} \geq 2 \mid z_s)$ can be estimated in redshift bins as
\begin{align}
    p\left(N_{\rm img} \geq 2 \mid z_s\right) \approx
    \frac{N_b}{\sum_{i \in b} N_{{\rm draw},\,i}},
    \quad z_s \in \mathcal{Z}_b ,
    \label{eq:acceptance_prob}
\end{align}
where $\{\mathcal{Z}_b\}$ are the uniform bins partitioning
$[z_{\rm min}, z_{\rm max}]$, $\mathcal{Z}_b$ is the bin containing $z_s$,
$N_b$ is the number of accepted systems with source redshift in
$\mathcal{Z}_b$, $N_{{\rm draw},\,i}$ is the number of proposal draws
required to accept the $i$-th such system, and the sum runs over those
$N_b$ systems.

With this in hand, we can approximate
Eq.~\ref{eq:target_detection_rate} by assigning an importance weight to
each simulated glSLSN-I. The weight assigned to the $i$-th realization is
defined as the ratio between the integrand of
Eq.~\ref{eq:target_detection_rate},
$\mathcal{I}(z_{s,i}, \boldsymbol{\theta}_{{\rm sl},i})$, and the sampling
distribution given by Eq.~\ref{eq:sampling_dist},
\begin{align}
    w_i = &\frac{\Omega_{\rm survey}}{16\pi^2}\frac{N_{\rm gal} N_q N_\gamma V_{\rm C}(z_{s,i}) (z_{\rm max} - z_{\rm min})}{N_{\rm glSLSN}} \nonumber \\
    &\times \frac{r_{\rm SLSN}(z_{s,i})}{1+z_{s,i}} \left.\frac{dV_{\rm C}}{dz}\right\rvert_{z=z_{s,i}} \nonumber \\
    &\times A_{{\rm box},i}\, p\left(N_{\rm img} \geq 2 \mid z_{s,i}\right) \epsilon_{{\rm det},i},
    \label{eq:importance_weight}
\end{align}
where $\Omega_{\rm survey}$ is the effective sky area of the survey under
consideration (Sec.~\ref{sec:ztf_survey}), and
$\epsilon_{{\rm det},i} = \epsilon_{\rm det}(m_{{\rm peak},i})$
(Eq.~\ref{eq:det_eff}) for realizations passing the photometric detection
criteria of Sec.~\ref{sec:glslsne_simulation}, with $w_i = 0$ otherwise.
The detection rate is then estimated as
$\dot{N}_{\rm det} \approx \sum_i w_i$.

The uncertainties quoted on all simulated detection rates propagate two
sources of error: the posterior uncertainty of the volumetric-rate model
and the finite size of the simulated catalog. For each posterior draw of Eq.~\ref{eq:slsne_rate}, we recompute the importance weights of Eq.~\ref{eq:importance_weight} and accumulate the detection rate in bins of source redshift. Within each bin, the Monte Carlo error of the weighted estimator is included by replacing the accumulated rate $\sum_i w_i$ with $\bar{w}\,n$, where 
\begin{align}
    &n \sim \mathrm{Poisson}(N_{\rm eff}),\nonumber\\
    &N_{\rm eff} = \left(\sum_i w_i\right)^{2} / \sum_i w_i^{2},
\end{align}
is the effective number of contributing systems, and $\bar{w}= \sum_i w_i^{2} / \sum_i w_i$; this reproduces the mean and, approximately, the variance of the importance-sampling estimator. Quoted rates and their intervals are the medians and 16th--84th percentiles, and are dominated by the volumetric-rate posterior, with the Monte Carlo contribution subdominant.

\section{Detection Rates}
\label{sec:detection_rates}

\begin{figure*}[ht!]
    \plotone{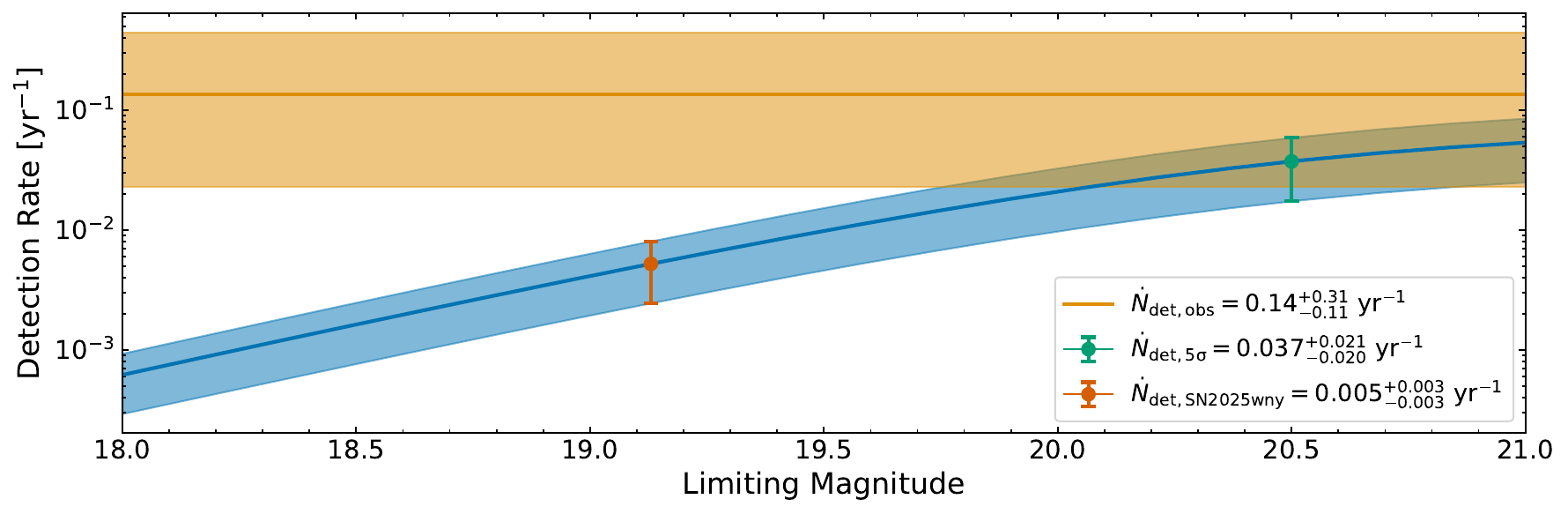}
    \caption{Simulated ZTF detection rate of glSLSNe-I as a function of the
    limiting magnitude $m_{\rm lim}$, defined as the peak apparent magnitude at
    which the spectroscopic selection efficiency of Eq.~\ref{eq:det_eff} reaches
    $50\%$. The solid blue line shows the median simulated detection rate and the
    blue band its $68\%$ confidence interval. The solid orange line and band show
    the observed detection rate implied by one discovery in seven years of ZTF
    operations, $\dot{N}_{\rm det,obs}=0.14^{+0.31}_{-0.11}$~yr$^{-1}$. Markers
    denote the two reference selection functions discussed in
    Sec.~\ref{sec:detection_rates}: the fiducial choice anchored to the ZTF
    single-visit $5\sigma$ depth, $m_{\rm lim}=20.5$, with
    $\dot{N}_{\rm det,5\sigma}=0.037^{+0.022}_{-0.020}$~yr$^{-1}$ (green), and the
    choice anchored to the peak apparent magnitude of SN~2025wny,
    $m_{\rm lim}=19.13$, with
    $\dot{N}_{\rm det,wny}=0.005^{+0.003}_{-0.003}$~yr$^{-1}$ (vermillion), with
    error bars showing the corresponding $68\%$ intervals.}
    \label{fig:glslsne_det_rates}
\end{figure*}

SN~2025wny is, to date, the only glSLSN-I discovered during seven years of ZTF
operations. Treating discovery as a Poisson process, a single event over this
period corresponds to an observed detection rate of
\begin{align}
    \dot{N}_{\rm det,obs} = 0.14^{+0.31}_{-0.11}~{\rm yr}^{-1},
\end{align}
where the uncertainties denote the $68\%$ confidence interval for a Poisson
process with one observed count. As
expected for a single event, the resulting interval spans more than an order of
magnitude, and any comparison with the simulated rate is correspondingly
coarse. Note that while the observed rate is derived from the full seven-year
baseline, the simulated rates below are computed from 2.7~yr of survey logs
(Sec.~\ref{sec:ztf_survey}) and annualized; the comparison therefore assumes
that the survey properties over the logged period are representative of ZTF
operations as a whole.

Comparing the observed rate with the simulations requires a model for the probability that a photometrically detected glSLSN-I is also identified as lensed and spectroscopically classified. For BTS and Z+LIONS this is well characterised (Eq. \ref{eq:det_eff}); the targeted search that found SN~2025wny has no comparably simple description, depending on lens-catalogue completeness as well as on follow-up depth. Following Sec. \ref{sec:ztf_survey} we set $\epsilon_{\rm flag}=1$ and assume a magnitude-dependent spectroscopic follow-up selection function. Rather than adopting a single, necessarily arbitrary, choice, we present the simulated detection
rate as a function of the limiting magnitude $m_{\rm lim}$, defined as the
peak apparent magnitude at which the spectroscopic efficiency of
Eq.~\ref{eq:det_eff} reaches $50\%$ (i.e.\ $m_0 = m_{\rm lim}$, with the slope
fixed to $s = 4.5$). Figure~\ref{fig:glslsne_det_rates} shows the resulting
median simulated rate and its $68\%$ confidence interval alongside the observed
rate. The simulated rate rises steeply between $m_{\rm lim} \simeq 19$ and
$21$, reflecting the faint peak apparent magnitudes of the underlying glSLSNe-I
population, before the growth slows as the spectroscopic limit approaches the photometric detection floor set by the ZTF single-visit depth.

As a fiducial case in the analysis of the targeted glSN search, we adopt $m_{\rm lim} = 20.5$, equal to the $5\sigma$
single-visit depth of ZTF, such that the $50\%$ spectroscopic-efficiency threshold coincides with the photometric detection floor and events
more than $\sim 0.5$~mag brighter than the floor are identified with
near-unit efficiency. Since any realistic spectroscopic program loses
efficiency before reaching the photometric limit, this choice serves as an
optimistic benchmark for the achievable detection rate. Under this assumption we find
\begin{align}
    \dot{N}_{\rm det,5\sigma} = 0.037^{+0.022}_{-0.020}~{\rm yr}^{-1},
\end{align}
in agreement with the observed rate within the mutual $1\sigma$ intervals.
Since the true efficiency of a manually driven search is necessarily below this
idealized limit, the agreement should be read as consistency between the upper
bound and the (weakly constrained) observed rate rather than as a precise
validation. If we instead anchor the selection function to SN~2025wny itself,
setting $m_{\rm lim} = 19.13$ such that the spectroscopic efficiency is $50\%$
at its peak apparent magnitude in ZTF, the predicted rate drops to
$\dot{N}_{\rm det,wny} = 0.005^{+0.003}_{-0.003}~{\rm yr}^{-1}$. At this rate
the probability of at least one detection over the seven-year baseline is only
$\approx 3$--$5\%$, placing this conservative choice in mild ($\sim 2\sigma$)
tension with the observed rate and suggesting that the effective depth of the
manual search exceeds the peak apparent magnitude of SN~2025wny. The two
choices together bracket the plausible range of the true, unmodeled selection
function.

For the untargeted ZTF surveys, whose selection functions are explicitly
modeled by Eq.~\ref{eq:det_eff}, we predict
\begin{align}
    \dot{N}_{\rm det,BTS} &= 0.003^{+0.002}_{-0.002}~{\rm yr}^{-1}, \\
    \dot{N}_{\rm det,Z+LIONS} &= 0.028^{+0.017}_{-0.015}~{\rm yr}^{-1}.
\end{align}
The shallow BTS limit of $m_{\rm lim} = 18.8$ excludes the bulk of the glSLSNe-I
population, which peaks well below this magnitude, as reflected in the steep
rise of the simulated rate with $m_{\rm lim}$ in
Fig.~\ref{fig:glslsne_det_rates}, yielding an expected detection rate of
roughly one event per $\sim 300$ years; the discovery of SN~2025wny through
BTS-like channels alone would thus have been highly unlikely. Z+LIONS, in contrast, reaches
$m_{\rm lim} = 20.6$ and recovers a rate comparable to the idealized full-survey upper
bound, $\dot{N}_{\rm det,Z+LIONS} \simeq \dot{N}_{\rm det,5\sigma}$, despite
covering only $10{,}000~{\rm deg}^2$ compared to the assumed BTS sky area of
$14{,}400~{\rm deg}^2$: the gain in depth partially compensates for the
reduced footprint. At this rate, Z+LIONS should yield $0.28^{+0.17}_{-0.15}$ glSLSNe-I over the 10 year duration of LSST operations.

The uncertainties on the simulated rates, constructed as described in Sec.~\ref{sec:importance_sampling}, are dominated by the poorly
constrained volumetric rate of SLSNe-I. These uncertainties are, moreover, underestimated:
we have not propagated systematic uncertainty in the assumed luminosity
function, which is derived from only 33 low-redshift events
\citep{chen_hydrogen-poor_2023}, nor in the magnetar-driven template grid
\citep{kessler_models_2019}. The consistency between the observed and
simulated rates should therefore be interpreted as a successful order-of-magnitude
test of the population and selection model, complementary to the
event-level consistency check provided by the debiasing analysis of
Sec.~\ref{sec:debiasing}.

\section{Debiasing of SN~2025wny}
\label{sec:debiasing}

Observations of glSNe such as SN~2025wny are impacted by typical observational
biases such as Malmquist bias, akin to other transients, but are additionally
affected by magnification bias. Magnification bias describes the tendency of
detected events to be biased towards higher magnifications, as these result in
brighter events, and becomes increasingly impactful as sources are shifted to
higher redshifts. Not taking such biases into account impacts astrophysically relevant properties of the observed transient, such as its
intrinsic peak bolometric luminosity. The intrinsic luminosity is of
particular interest for SN~2025wny, as it directly informs whether the event is
consistent with the known SLSNe-I population or represents an intrinsically
extreme event.

These biases can be accounted for via simulation of the observed system,
assuming known SED templates and an assumed luminosity function. Compared to
transients such as SNe Ia, however, neither is well constrained for SLSNe-I;
we return to this caveat below. With this in mind, we debias the
magnification posterior presented in \cite{Mortsell2026} and infer the debiased
intrinsic peak bolometric luminosity under the luminosity function and
template distribution adopted earlier in this paper. Due to the mismatch
between the assumed deflector population and the actual lensing system
described in \cite{Mortsell2026}, we re-simulate $N = 5\times10^{7}$ glSLSNe-I
according to the pipeline described in Sec.~\ref{sec:glslsne_simulation}, with
two changes: we replace the deflector population with the
lens-model magnification posterior of \cite{Mortsell2026}, sampled jointly across
images such that inter-image correlations are preserved, and we fix the source
redshift to $z_s = 2.015$.


\begin{figure}[ht!]
    \includegraphics[width=\linewidth]{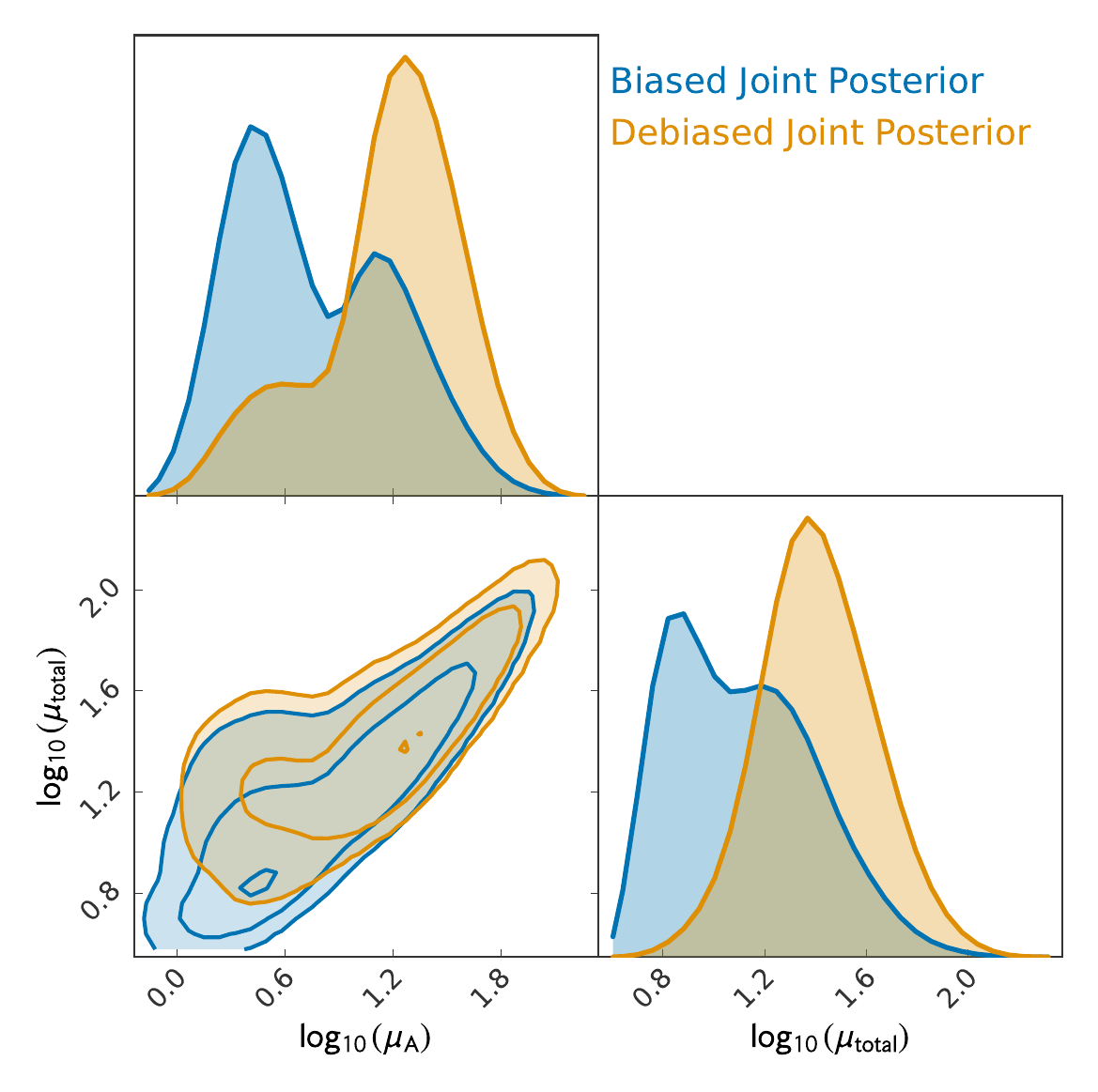}
    \caption{Marginalized joint posterior distributions of the image-A and total magnifications. Blue and orange denote the biased (Equation \ref{eq:biased_posterior}) and debiased (Equation \ref{eq:debiased_posterior}) inferences, respectively, both using the observed luminosity of image D. One-dimensional marginal distributions are shown on the diagonal. The combination of the luminosity function and the ZTF selection function suppresses the low-magnification mode associated with image A, shifting both $\mu_{\rm A}$ and $\mu_{\rm tot}$ towards higher values. Summary statistics, including those for $\mu_{\rm }$ and $L_{\rm bol, int}$, which are not shown here, are provided in Table \ref{tab:posteriors}.}
    \label{fig:joint_posteriors}
\end{figure}

Each re-simulated system consists of an intrinsic peak bolometric luminosity
drawn from the luminosity function of Sec.~\ref{sec:slsne_population}, a
magnification vector $\boldsymbol{\mu} = (\mu_{\rm A}, \ldots, \mu_{\rm E})$
drawn from the lens-model posterior, and a template, Milky Way sightline, and
ZTF noise realization as described in
Secs.~\ref{sec:slsne_population}--\ref{sec:ztf_survey}. Each system is
assigned the spectroscopic selection weight $\epsilon_{\rm det}(m_{\rm peak})$
of Eq.~\ref{eq:det_eff} with $m_0 = 20.5$, evaluated at the peak apparent
magnitude of the unresolved lensed light curve. The weighted samples represent
the joint distribution of intrinsic luminosity and magnification for lensed
SLSNe-I in this lens system \emph{conditioned on discovery by ZTF},
\begin{align}
    p\left(\log_{10} L_{\rm bol,int}, \boldsymbol{\mu} \mid \mathrm{det}\right)
    \propto\;
    &p\left(\log_{10} L_{\rm bol,int}\right)\,
    p_{\rm lens}\!\left(\boldsymbol{\mu}\right) \nonumber \\
    &\times\,
    \bar{\epsilon}_{\rm det}\!\left(\log_{10} L_{\rm bol,int}, \mu_{\rm tot}\right),
    \label{eq:sim_prior}
\end{align}
where $p\left(\log_{10} L_{\rm bol,int}\right)$ is the luminosity function,
$p_{\rm lens}$ is the lens-model posterior of \cite{Mortsell2026}, and
$\bar{\epsilon}_{\rm det}$ is the selection probability marginalized over
templates, sightlines, and noise realizations. Since discovery depends on the
brightness of the unresolved light curve, $\bar{\epsilon}_{\rm det}$ depends
on the magnification vector primarily through the total magnification
$\mu_{\rm tot}$. Equation~\ref{eq:sim_prior} makes the origin of the
magnification bias explicit: conditioned on detection, the intrinsic
luminosity and the magnification are anti-correlated, since intrinsically
faint sources only enter the detected population when highly magnified.

Given a measurement of the magnified peak bolometric luminosity of a single image $X$, the lens-only, or biased, inference of the intrinsic luminosity can be expressed as the posterior
\begin{align}
    &p_{\rm bias}\left(\log_{10} L_{\rm bol,int}, \boldsymbol{\mu} \mid L_{\rm obs, X}\right) \nonumber \\
    &\propto 
    \mathcal{L}_{\rm X}\!\left(
        \log_{10} L_{\rm bol,int} + \log_{10}\mu_{\rm X}
    \right)
    p_{\rm lens}\!\left(\boldsymbol{\mu}\right)
    \pi(\log_{10} L_{\rm bol,int}) 
    \label{eq:biased_posterior}
\end{align}
where $\mathcal{L}_{\rm X}$ is the likelihood of the measured luminosity given the
true magnified luminosity of image $X$, $p_{\rm lens}$ is the lens-model posterior, and $\pi$ denotes the assumed prior density in intrinsic luminosity $L_{\rm bol,int}$, which is assumed to be flat. 
Neither the flat prior assumption nor the lens-model posterior as a lens prior assumption holds for a magnitude-limited discovery: the resulting estimate ignores both the shape of the luminosity function and the
selection-induced correlation of Eq.~\ref{eq:sim_prior}, biasing the recovered intrinsic luminosity.

The debiased posterior is instead obtained by treating Eq.~\ref{eq:sim_prior}
as the prior and updating it with the observed luminosity,
\begin{align}
    p&\left(\log_{10} L_{\rm bol,int}, \boldsymbol{\mu}
    \mid L_{\rm obs, X}, \mathrm{det}\right)
    \propto \nonumber \\
    &\mathcal{L}_{\rm X}\!\left(\log_{10} L_{\rm bol,int} + \log_{10}\mu_{\rm X}\right)
    p\left(\log_{10} L_{\rm bol,int}, \boldsymbol{\mu} \mid \mathrm{det}\right),
    \label{eq:debiased_posterior}
\end{align}
In practice this update is performed by importance
reweighting: each simulated system is assigned the weight
$\tilde{w}_k = w_k\,
\mathcal{L}_{\rm X}(\log_{10} L_{{\rm bol,int},k} + \log_{10}\mu_{{\rm X},k})$, where
$w_k$ is its selection weight, and $\mathcal{L}_{\rm X}$ is estimated via a Gaussian
kernel density estimate over Monte Carlo realizations of the measured
luminosity. The reweighted samples then represent the debiased joint posterior
over $(\log_{10} L_{\rm bol,int}, \boldsymbol{\mu})$. We note that the update in
Eq.~\ref{eq:debiased_posterior} acts on the full magnification vector: through
the luminosity function and the selection term, the observed luminosity of a
single image informs the magnifications of all images. In this sense the
transient itself acts as an additional constraint on the lens model, analogous
to the use of lensed SNe Ia as standardizable sources, but with the standard
candle replaced by a broad luminosity function. This is also the central
caveat of the method: the debiased posterior is conditional on the assumed
luminosity function and template distribution, which for SLSNe-I are derived
from a small spectroscopically-complete sample \citep{chen_hydrogen-poor_2023} and a magnetar-driven model grid \citep{kessler_models_2019}, respectively, and
carry systematic uncertainties that are difficult to quantify. The resulting
constraints should therefore be interpreted as conditional on this population
model, rather than as fully model-independent measurements.

The observed pseudo-bolometric light curve of ~SN~2025wny is constructed from photometry of image A \citep{Li2026}. However, microlensing broadens the magnification posterior of image A, whereas image D is more tightly constrained by the lens model \citep{Mortsell2026}. We therefore follow the image-D-based de-lensing procedure of \cite{Li2026}, setting $X=D$ in Equations \ref{eq:biased_posterior} and \ref{eq:debiased_posterior}. We convert the image-A-based observed luminosity according to

\begin{align}
    L_{{\rm obs,}D} = F_{D/A}L_{{\rm obs,}A},
\end{align}
adopting their phase-corrected flux ratio $F_{D/A}=0.160\pm 0.008$. This ratio is obtained by combining Gaussian Process (GP) interpolation of the ground-based image-A light curves with resolved HST photometry of image D, accounting for the measured time delay. Its uncertainty includes the GP, time-delay, and photometric contributions; we refer to Section 3.1 of \cite{Li2026} for details.

\begin{table}
\begin{tabular}{lcc}
\hline
 & Biased & Debiased \\
\hline
$\mu_{\rm A}$ 
& $4.6_{-2.5}^{+13.2}$ 
& $17.1_{-11.6}^{+17.8}$ \\

$\mu_{\rm D}$ 
& $3.09_{-0.27}^{+0.60}$ 
& $3.13_{-0.28}^{+0.76}$ \\

$\mu_{\rm total}$ 
& $12.0_{-5.5}^{+13.2}$ 
& $25.1_{-9.3}^{+17.7}$ \\

$\log_{10}(L_{\rm bol,int}/{\rm erg\,s^{-1}})$ 
& $44.60_{-0.08}^{+0.05}$ 
& $44.60_{-0.10}^{+0.05}$ \\
\hline
\end{tabular}
\caption{Medians and $16$th-$84$th percentile intervals of the biased (Equation \ref{eq:biased_posterior}) and debiased (Equation \ref{eq:debiased_posterior}) marginalized posterior distributions for ~SN~2025wny. Figure \ref{fig:joint_posteriors} shows the corresponding distributions for the image-A and total magnifications.}
\label{tab:posteriors}
\end{table}


\begin{figure*}[ht!]
    \plotone{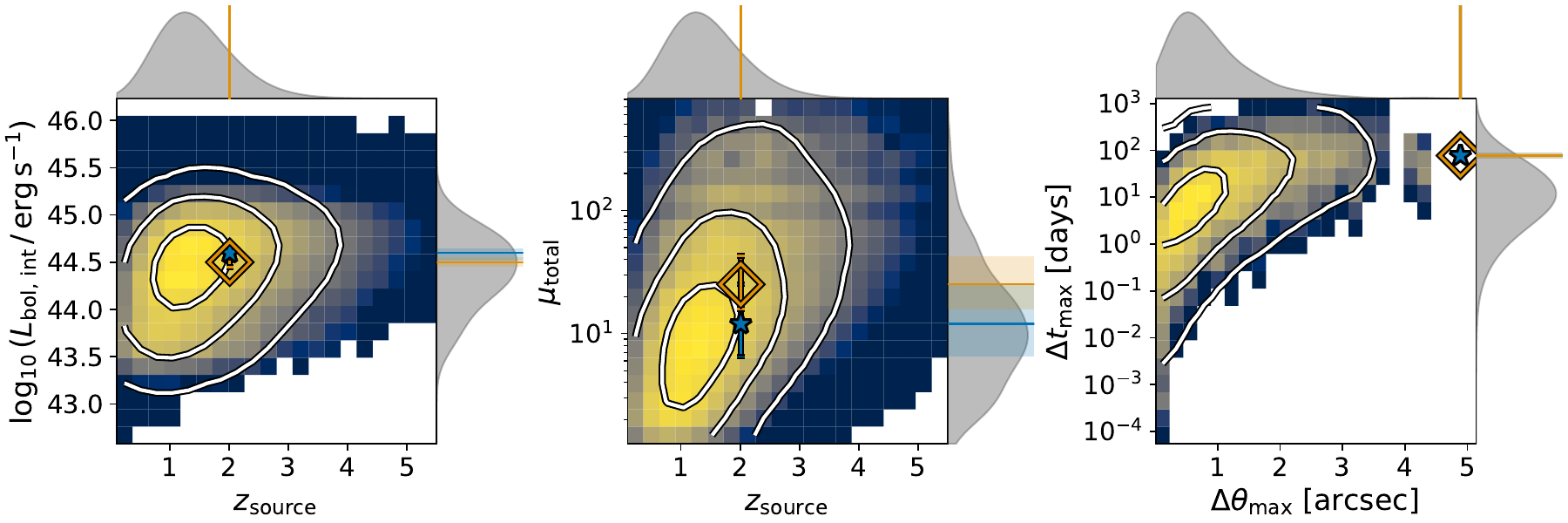}
    \caption{Joint distributions of the properties of ZTF-detected
    glSLSNe-I, combining doubly, triply, and quadruply imaged systems and weighted by the importance weights of
    Sec.~\ref{sec:importance_sampling} under the fiducial selection function
    with $m_{\rm lim}=20.5$ (Sec.~\ref{sec:detection_rates}). Panels show,
    from left to right, the joint rate density distributions of
    $(z_s, \log_{10} L_{\rm bol,int})$, $(z_s, \mu_{\rm tot})$,
    and $(\Delta\theta_{\rm max}, \Delta t_{\rm max})$, where
    $\Delta\theta_{\rm max}$ is the maximum image separation and
    $\Delta t_{\rm max}$ the maximum time delay. Marginal distributions are shown in grey. White contours enclose the
    $39.3\%$-, $86.5\%$-, and $98.9\%$ regions corresponding to $1$-, $2$- and $3\sigma$. Markers denote the corresponding values for SN~2025wny:
    the spectroscopic time delay is taken from
    \cite{Johansson2026}, the maximum image separation is taken from \cite{Goobar2026}, while the magnification and intrinsic luminosity are the biased (blue star) and debiased (orange diamond) posteriors of
    Sec.~\ref{sec:debiasing}, with error bars and shaded bands denoting the
    $68\%$ intervals.}
    \label{fig:props}
\end{figure*}

Figure~\ref{fig:joint_posteriors} compares the biased
(Eq.~\ref{eq:biased_posterior}) and debiased (Eq.~\ref{eq:debiased_posterior})
joint posteriors marginalized onto $\mu_{\rm A}$ and $\mu_{\rm tot}$. Table \ref{tab:posteriors} reports the medians and $16$th-$84th$ percentile intervals for $\mu_{\rm A}$, $\mu_{\rm D}$, $\mu_{\rm tot}$, and $L_{\rm bol, int}$. Both inferences use identical lens-model samples and the same observed-luminosity likelihood, so their differences reflect the adopted luminosity function and selection function.

The marginalized posteriors for $\mu_{\rm D}$ and $L_{\rm bol, int}$, which are not shown in Figure~\ref{fig:joint_posteriors}, change only marginally (Table \ref{tab:posteriors}). In particular, both intrinsic-luminosity posteriors have a median of $\log_{10}(L_{\rm bol, int}/{\rm erg\, s^{-1}})=44.60$, with a modestly extended faint tail after debiasing. The dominant effect
of the update is instead on image A: with the intrinsic luminosity pinned by
the likelihood and the luminosity function, discovery of a $z_s = 2.015$
SLSN-I by ZTF requires a large total magnification, and the selection term
$\bar{\epsilon}_{\rm det}$ therefore strongly favors the high-magnification
mode of the bimodal $\mu_{\rm A}$ posterior. The debiased posteriors on
$\mu_{\rm A}$ and $\mu_{\rm tot}$ are correspondingly shifted towards higher
magnifications and bimodalities are suppressed. The population-level
selection model thus partially resolves a degeneracy that the lens modeling alone,
affected by microlensing of image A, could not.

The near-agreement between the biased and debiased intrinsic-luminosity
posteriors admits two complementary readings. First, it is a robustness
statement about the measurement: because $\mu_{\rm D}$ is tightly constrained
by the lens model, the inferred $L_{\rm bol,int}$ leans only weakly on the
population model and is therefore largely insensitive to the systematic
uncertainties in the assumed luminosity function and templates discussed
above. The debiased $\mu_{\rm A}$ and $\mu_{\rm tot}$ constraints, by
contrast, are driven by the selection term and inherit those systematics in
full. Second, it is a non-trivial consistency check of the population model
itself: the intrinsic luminosity implied by the well-constrained image-D
magnification falls within the bright tail of the assumed luminosity function, at
the $96.9{\rm th}^{+0.9}_{-1.8}$ percentile, as shown in Fig.~\ref{fig:luminosity}. Had the assumed luminosity function been strongly inconsistent with SN~2025wny, the debiased posterior would instead
have been displaced away from the naive estimate towards the bulk of the luminosity function. The agreement therefore indicates joint consistency between the lens model,
the luminosity function, and the selection model, and implies that the
extreme apparent brightness of SN~2025wny is fully attributable to lensing
rather than to an intrinsically anomalous event. It does not, however,
constrain the shape of the luminosity function beyond the location of
SN~2025wny within it.

\section{glSLSNe-I Population Properties}
\label{sec:population_properties}

Figure~\ref{fig:props} presents the joint distributions of the properties of
ZTF-detectable glSLSNe-I, obtained by weighting each simulated system by its
importance weight (Sec.~\ref{sec:importance_sampling}) under the fiducial
selection function of Sec.~\ref{sec:detection_rates} ($m_{\rm lim} = 20.5$) and assuming the median SLSNe-I volumetric rate (Sec.~\ref{sec:slsne_population}).
From left to right, the panels show $(z_s, \, \log_{10}L_{\rm bol,int})$, $(z_s, \,\mu_{\rm tot})$ and $(\Delta\theta_{\rm max},\, \Delta t_{\rm max})$, where $\Delta\theta_{\rm max}$ and $\Delta t_{\rm max}$ are the maximum image separation and maximum time delay, respectively. Doubly, triply, and quadruply imaged systems exhibit broadly similar distributions of the properties considered here, so we present and discuss the combined ZTF-detectable population. For comparison, we overplot the corresponding quantities for SN~2025wny: the
spectroscopic time delay is taken from \cite{Johansson2026}, the maximum image separation from \cite{Goobar2026}, and the biased and debiased magnifications and intrinsic luminosities from Table \ref{tab:posteriors}.

The intrinsic properties of the detected population are shaped by the
interplay of two selection effects. The $(z_s, L_{\rm bol,int})$ panel shows
a clear Malmquist bias: as the source redshift increases, intrinsically
faint events fall below the detection threshold, and the detected population
is progressively restricted to the bright end of the luminosity function. The $(z_s, \mu_{\rm tot})$ panel shows the corresponding magnification bias:
the typical total magnification of detected systems rises steeply with redshift, since at high $z_s$ only strongly magnified events remain
observable. Together, these effects determine which combinations of intrinsic luminosity and magnification enter the detected population, with intrinsically fainter sources requiring greater magnification to remain observable.

In these intrinsic properties, SN~2025wny is consistent with being drawn from
the simulated ZTF-detectable population. In the $(z_s, L_{\rm bol,int})$ plane
the event lies on the boundary of the $1\sigma$ region. The biased $\mu_{\rm tot}$ lies on the $1\sigma$
boundary of the $(z_s, \mu_{\rm tot})$ plane, while the larger
debiased $\mu_{\rm tot}$ lies between the $1\sigma$ and $2\sigma$ contours. The upward shift in $\mu_{\rm tot}$
produced by the debiasing thus moves the event from the peak of the detected magnification distribution into its upper tail. This is not a sign of inconsistency but a consequence of the source redshift: the marginal magnification distribution is dominated by lower-redshift systems, which require little magnification to be detected, whereas at $z_s = 2.015$
detection demands a large total magnification, as seen in the
$(z_s, \mu_{\rm tot})$ panel. Conditioned on its redshift, the debiased magnification of SN~2025wny is fully in line with the detectable population,
and its extreme apparent brightness remains a consequence of lensing and
selection rather than of an intrinsically anomalous event.

The image-plane properties tell a different story. With a maximum image
separation of $\Delta\theta_{\rm max} \simeq 4.9''$ \citep{Goobar2026}, SN~2025wny
lies beyond the $3\sigma$ contours of the simulated population in the $(\Delta\theta_{\rm max},\, \Delta t_{\rm max})$ panel of Figure \ref{fig:props}. This is a direct
consequence of the assumed deflector population: our simulations model each
lens as a single SIE with external shear, for which image separations of
$\sim 5''$ require velocity dispersions in the exponentially suppressed tail
of the velocity-dispersion function (Eq.~\ref{eq:vd}). The actual lensing
configuration of SN~2025wny is more complex, consisting of two deflector
galaxies at the same redshift, potentially embedded in the central region of
a galaxy cluster \citep{Mortsell2026}. Such compound, group-assisted configurations
are absent from our model by construction. More realistic deflector
populations that include multi-galaxy deflectors and their host halos, such
as the halo-model-based mock catalogs of SL-Hammocks \citep{abe_halo_2025},
would likely capture this class of wide-separation, galaxy-scale lensing.
The discovery of SN~2025wny in precisely this regime suggests that our
simulations underestimate the rate of systems with large image separations
and, correspondingly, long time delays; the quoted detection rates of
Sec.~\ref{sec:detection_rates} should therefore be regarded as conservative
with respect to this population. We note that the observed time delay itself
is not anomalous: the $(\Delta\theta_{\rm max}, \Delta t_{\rm max})$ panel
shows that SN~2025wny follows the extrapolation of the strong correlation
between image separation and time delay traced by the simulated population,
so it is the separation, not the delay at fixed separation, that places the
event outside the model.

Finally, we recall that the simulated population neglects microlensing by
stars in the deflector (Sec.~\ref{sec:unresolved_lightcurves}). Its expected
population-level effect is a broadening of the magnification, and hence
inferred luminosity, distributions, rather than a systematic shift in the
detection rates \citep{arendse_detecting_2024}. Given that image A of
SN~2025wny is itself significantly microlensed
(Sec.~\ref{sec:debiasing}), the contours of Fig.~\ref{fig:props} involving
$\mu_{\rm tot}$ and $L_{\rm bol,int}$ should be read as slightly narrower
than those of a fully realistic population.

\section{Summary and Conclusions}
\label{sec:summary}

The discovery of SN~2025wny, the first strongly lensed SLSN-I, raises two
population-level questions: whether a single such event in seven years of ZTF
operations is consistent with expectations for the known SLSNe-I population and
the ZTF selection function, and whether SN~2025wny itself is a typical
representative of the ZTF-detectable glSLSNe-I population. To address these
questions, we have developed a forward simulation of strongly lensed SLSNe-I
in ZTF, combining an empirically calibrated volumetric rate
(Eq.~\ref{eq:slsne_rate}, fit to literature measurements assuming a
CSFRD-driven parametrization) and a $V_{\rm max}$-weighted luminosity function
constructed from the spectroscopically complete $m_{\rm peak} < 19$ ZTF subsample of \citet{chen_hydrogen-poor_2023}, with a galaxy-scale SIE+shear deflector
population, unresolved lensed light curves including Milky Way extinction and
IGM attenuation, and the actual ZTF observing history over 2.7~yr of survey
logs. Importance-sampling weights (Sec.~\ref{sec:importance_sampling}) convert
the simulated catalog of $10^9$ lensed realizations into physical detection
rates. Our main conclusions are as follows.

\begin{enumerate}

\item \emph{The observed glSLSN-I detection rate in ZTF is consistent with
expectations.} Under the fiducial selection function, in which photometrically detected glSLSNe-I with a peak apparent magnitude equal to the ZTF single-visit $5\sigma$ depth $m_{\rm lim} = 20.5$ have a $50\%$ likelihood of spectroscopic identification, we predict $\dot{N}_{\rm det,5\sigma} = 0.037^{+0.022}_{-0.020}$~yr$^{-1}$, while anchoring the
selection function to the peak apparent magnitude of SN~2025wny itself
($m_{\rm lim} = 19.13$) yields
$\dot{N}_{\rm det,wny} = 0.005^{+0.003}_{-0.003}$~yr$^{-1}$. The fiducial estimate is consistent within the mutual $1\sigma$ intervals with
the observed rate of
$\dot{N}_{\rm det,obs} = 0.14^{+0.31}_{-0.11}$~yr$^{-1}$ implied by one
discovery in seven years, while the SN~2025wny-anchored estimate is in mild
($\sim 2\sigma$) tension with it, suggesting that the effective depth of the
manual search exceeds the peak apparent magnitude of SN~2025wny; together the
two choices bracket the plausible range of the true, unmodeled selection
efficiency of the current manually driven glSNe search. The discovery of
SN~2025wny is therefore unsurprising at the order-of-magnitude level, given
the known SLSNe-I population.

\item \emph{Untargeted spectroscopic surveys differ sharply in their
sensitivity to glSLSNe-I.} The shallow BTS limit excludes the bulk of the
faint glSLSNe-I population, yielding
$\dot{N}_{\rm det,BTS} = 0.003^{+0.002}_{-0.002}$~yr$^{-1}$, roughly one event
per $\sim 300$ years. The deeper Z+LIONS survey recovers
$\dot{N}_{\rm det,Z+LIONS} = 0.028^{+0.017}_{-0.015}$~yr$^{-1}$, comparable to
the idealized full-survey upper bound despite its smaller footprint,
corresponding to $0.28^{+0.17}_{-0.15}$ detected glSLSNe-I over the full 10 years of LSST operations.

\item \emph{The extreme apparent brightness of SN~2025wny is fully
attributable to lensing.} Conditioning the simulated population on discovery
by ZTF and updating with the observed image-D luminosity yields debiased
posteriors on the magnifications and intrinsic peak bolometric luminosity of
SN~2025wny (Sec.~\ref{sec:debiasing}). Because the macro-magnification of
image D is tightly constrained by the lens model, the debiased intrinsic
luminosity,
$\log_{10}(L_{\rm bol,int}/{\rm erg\,s^{-1}}) = 44.60^{+0.05}_{-0.10}$, is
nearly identical to the biased estimate and leans only weakly on the assumed
population model. The luminosity places SN~2025wny in the bright tail of the
SLSNe-I luminosity function, at the $96.9^{+0.9}_{-1.8}$ percentile: luminous,
but not anomalous. We stress that this statement concerns the peak luminosity
alone; the spectroscopic and photometric peculiarities of SN~2025wny relative
to the known low-redshift SLSNe-I population are discussed in \cite{Li2026}.

\item \emph{Selection modeling sharpens the lens-model magnifications.} The
combination of the luminosity function and the ZTF selection function
suppresses the low-magnification mode of the microlensing-induced bimodal
image-A magnification posterior, shifting the debiased constraints to
$\mu_{\rm A} = 17.1^{+17.8}_{-11.6}$ and
$\mu_{\rm tot} = 25.1^{+17.7}_{-9.3}$, compared to the biased values of
$4.6^{+13.2}_{-2.5}$ and $12.0^{+13.2}_{-5.5}$, respectively. The transient
itself thus acts as an additional constraint on the lens model, analogous to
the use of lensed SNe Ia as standardizable sources but with the standard
candle replaced by a broad luminosity function; these constraints are
correspondingly conditional on the assumed luminosity function and template
distribution.

\item \emph{SN~2025wny is typical of the detectable population in its
intrinsic luminosity and magnification, but geometrically
exceptional.} In source redshift, intrinsic luminosity, and total
magnification, SN~2025wny lies within the $1$--$2\sigma$ regions of the
simulated ZTF-detectable population; the debiased magnification places it in
the upper tail of the detected magnification distribution, as expected for a
source at $z_s = 2.015$, for which detection requires large total
magnification (Sec.~\ref{sec:population_properties}). In contrast, its maximum image
separation of $\simeq 4.9''$ places it beyond the $3\sigma$ contours of the simulated $(\Delta\theta_{\rm max},\, \Delta t_{\rm max})$ distribution. This reflects
a limitation of the assumed deflector model: single SIE deflectors reach such
separations only in the exponentially suppressed tail of the
velocity-dispersion function, whereas SN~2025wny is lensed by two galaxies,
potentially embedded in a group- or cluster-scale halo. The observed time
delay, however, follows the separation--delay correlation traced by the
simulated population; it is the separation itself that is anomalous.

\end{enumerate}

The chief limitations of this work follow directly from the above. The
predicted rates inherit the poorly constrained SLSNe-I volumetric rate in
full, and do not propagate systematic uncertainty in the luminosity function,
derived from only 33 low-redshift events, or in the magnetar-driven template
grid. The absence of compound and group-assisted deflectors implies that the
quoted rates are conservative with respect to wide-separation,
long-time-delay systems, which is precisely the regime in which SN~2025wny
was found and which is the most valuable for time-delay cosmography. Finally,
the neglect of microlensing modestly narrows the predicted magnification and
luminosity distributions.

These limitations point to clear directions for future work: replacing the
single-galaxy deflector population with halo-based mock catalogs that include
multi-galaxy deflectors and their host halos \citep{abe_halo_2025}, propagating
luminosity-function and template systematics into the rate predictions, and
extending the framework to the Rubin LSST era, where the increased depth
demonstrated by the Z+LIONS projection suggests that samples of glSLSNe-I,
with their long, time-dilated light curves and wide separations well matched
to time-delay cosmography, may finally become attainable. SN~2025wny, an intrinsically luminous but not anomalous SLSN-I, spectroscopically
distinct from the known low-redshift population \citep{Li2026}, placed in an
extraordinary lensing configuration, offers a first glimpse of that
population.

\begin{acknowledgments}

E.M.\ acknowledges support from the Swedish Research Council under Dnr VR 2024-03927. 

A.G.\ acknowledges financial support from the research project grant “Understanding the Dynamic Universe” funded by the Knut and Alice Wallenberg under Dnr KAW 2018.0067, {\em Vetenskapsr\aa det}, the Swedish Research Council through grants projects Dnr 2020-03444 and 2025-03692 the G.R.E.A.T research environment, Dnr 2016-06012, the EDUCATE excellence center funded by the Swedish Research Council through grant Dnr 2022-06627 and the Swedish National Space Agency, Dnr 2023-00226.

A. S. acknowledges support from the Knut and Alice Wallenberg Foundation through the ``Gravity Meets Light" project. 

A.T. and S.D acknowledge support from UK Research and Innovation (UKRI) under the UK government’s Horizon Europe funding Guarantee EP/Z000475/1.

E.E.H.\ is supported by a Gates Cambridge Scholarship (\#OPP1144).

C.V. acknowledges INAF project Supporto Arizona \& Italia.

\end{acknowledgments}



\bibliography{references}{}
\bibliographystyle{aasjournalv7}

\end{document}